\documentclass{article}

\usepackage[dblblindworkshop, final]{neurips_2025}
\workshoptitle{Scaling Environments for Agents (SEA)}

\usepackage[utf8]{inputenc} 
\usepackage[T1]{fontenc}    
\usepackage{hyperref}       
\usepackage{url}            
\usepackage{booktabs}       
\usepackage{amsmath}        
\usepackage{amsfonts}       
\usepackage{nicefrac}       
\usepackage{microtype}      
\usepackage{xcolor}         
\usepackage{graphicx}       
\usepackage{caption}        

\title{Agentic Persona Control and Task State Tracking for Realistic User Simulation in Interactive Scenarios}

\author{
  \textbf{Hareeshwar Karthikeyan} \\
  \textbf{Data Scientist} \\
  \textbf{Toast Inc.} \\
  \textbf{\texttt{hareesh.karthik@toasttab.com}} \\
}

\begin{document}

\maketitle

\begin{abstract}
Testing conversational AI systems at scale across diverse domains necessitates realistic and diverse user interactions capturing a wide array of behavioral patterns. We present a novel multi-agent framework for realistic, explainable human user simulation in interactive scenarios, using persona control and task state tracking to mirror human cognitive processes during goal-oriented conversations. Our system employs three specialized AI agents: (1)~a~User Agent to orchestrate the overall interaction, (2)~a~State Tracking Agent to maintain structured task state, and (3)~a~Message Attributes Generation Agent that controls conversational attributes based on task progress and assigned persona. To validate our approach, we implement and evaluate the framework for guest ordering at a restaurant with scenarios rich in task complexity, behavioral diversity, and conversational ambiguity. Through systematic ablations, we evaluate the contributory efficacy of each agentic component to overall simulation quality in terms of persona adherence, task completion accuracy, explainability, and realism. Our experiments demonstrate that the complete multi-agent system achieves superior simulation quality compared to single-LLM baselines, with significant gains across all evaluation metrics. This framework establishes a powerful environment for orchestrating agents to simulate human users with cognitive plausibility, decomposing the simulation into specialized sub-agents that reflect distinct aspects of human thought processes applicable across interactive domains.
\end{abstract}

\section{Introduction}

The rapid deployment of conversational AI systems across diverse customer-facing applications from restaurant ordering and e-commerce to healthcare consultations and customer support~\cite{Rastogi2019SchemaGuidedDialogue,Sun2022MetaphoricalUserSimulators} has created an urgent need for comprehensive testing methodologies that can simulate realistic human user behavior~\cite{Balog2025UserSimulationEraGenerativeAI}. Current approaches rely on static test sets or human evaluators, both presenting significant limitations~\cite{Davidson2023UserSimulationLLM,Ahmad2025SimulatingUserDiversity}. Static tests fail to capture the dynamic, multi-turn nature of human conversations, while human evaluation is expensive, difficult to scale, and challenging to standardize across different interaction scenarios~\cite{Davidson2023UserSimulationLLM,Zhuge2024AgentAsAJudge}. Moreover, existing automated testing approaches typically lack the behavioral diversity and contextual awareness necessary to simulate realistic user interactions~\cite{Ahmad2025SimulatingUserDiversity,Park2025SimulatingHumanBehavior}. Traditional single-model approaches struggle to balance these requirements, producing either overly scripted interactions that fail to adapt, or unpredictable behaviors that compromise reliability and evaluation consistency~\cite{Chu2024CohesiveConversations,Castricato2024PERSONA,Mehri2025GoalAlignment,Wang2025SurveyLLMAgents}.

\begin{figure}[h]
\centering
\includegraphics[width=\textwidth]{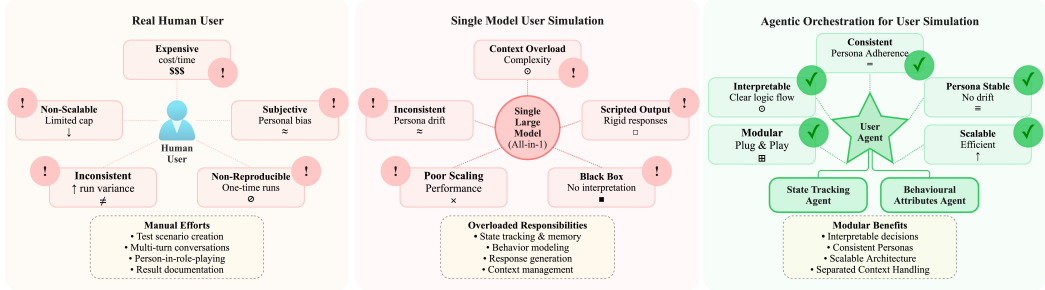}
\caption{\textbf{Comparison of simulation approaches:} \textbf{Left panel (Human User)}: Manual testing requires significant human effort including test scenario creation, multi-turn conversation management, person-in-role-playing, and result documentation, making it expensive and difficult to scale. \textbf{Center panel (Single Model System)}: Traditional automated approaches using a single model suffer from overloaded responsibilities, attempting to simultaneously handle state tracking and memory, behavior modeling, response generation, and context management, leading to inconsistent personas and poor interpretability. \textbf{Right panel (Agentic Simulation)}: Our proposed multi-agent framework distributes intelligence across specialized components, providing interpretable decisions, reproducible behavior, consistent personas, scalable architecture, and separated concerns, enabling systematic and reliable user simulation at scale. This decomposition mirrors human cognitive processes: tracking task completion progress (working memory)~\cite{Sun2022MetaphoricalUserSimulators,Hu2025UnifiedMindModel}, deciding how to respond based on personality and context (behavioral planning)~\cite{Park2023GenerativeAgents}, and generating appropriate utterances (language production).}
\label{fig:architecture}
\end{figure}

In this work, we propose a \textbf{multi-agent orchestration framework for human user simulation} in interactive scenarios that decomposes user behavior modeling into smaller, specialized components~\cite{Dang2025MultiAgentCollaboration,Lee2024OrchestraLLM}. Instead of relying on a single model, the framework employs distinct agents for managing task state, generating behavioral attributes, and coordinating interactions through structured protocols. To validate our approach, we implement and evaluate the framework in restaurant ordering - a domain that reflects the complexities of human interaction through multi-turn conversations, complex state tracking, and diverse persona-driven behaviors~\cite{Rastogi2019SchemaGuidedDialogue}. The framework rests on three core concepts: \textbf{Task State Management}, where a State Tracking Agent maintains a structured representation of the evolving task state, enabling precise progress tracking~\cite{Mehri2025GoalAlignment,Niu2024EnhancingDST}; \textbf{Behavioral Attribute Control}, where a Message Attributes Generation Agent dynamically determines conversational traits (mood, task execution style, exploration patterns) while preserving persona consistency~\cite{Ahmad2025SimulatingUserDiversity,WangChiu2023HumanoidAgents}; and \textbf{Tool-mediated Coordination}, where structured protocols govern agent interactions, ensuring proper context sharing without overlap of responsibilities~\cite{Raza2024TRiSM}. 

To our knowledge, this is the first work to explore explainable realistic human user simulation through a multi-agent architecture that combines dedicated agentic task tracking with fine-grained message generation attribute control. The unique internal environment we assess where specialized agents collaborate through structured protocols to maintain both task state coherence and persona consistency represents a novel approach in the user simulation landscape. This novelty informs our evaluation methodology, which focuses on demonstrating the framework's effectiveness through systematic ablation studies rather than direct comparisons with existing user simulation approaches that operate under fundamentally different architectural assumptions.

In summary, our work makes the following contributions:

\begin{itemize}
    \item[$\checkmark$] A novel multi-agent framework for human user simulation in interactive scenarios with specialized agents improving realism, controllability, and explainability through persona control and task state grounding
    
    \item[$\checkmark$] Systematic evaluation methodology with ablation studies and standardized metrics for persona adherence, task completion accuracy, decision explainability and overall simulation quality
    
    \item[$\checkmark$] Comprehensive test dataset in the restaurant ordering domain with 60 ordering test cases to validate the framework's effectiveness in complex, multi-turn conversational scenarios

\end{itemize}

\section{Related Works}

\paragraph{Human Simulation and Persona Modeling} AI agents demonstrate remarkable progress in simulating human behavior. Park et al.~\cite{Park2024GenerativeAgentSimulations,Park2025SimulatingHumanBehavior} show generative agents replicate survey responses with 85\% accuracy compared to human self-consistency, while Park et al.~\cite{Park2023GenerativeAgents} introduce architectures combining memory, reflection, and planning for believable behavior including emergent social interactions. Persona modeling has evolved from descriptive sentences~\cite{Sutcliffe2023SurveyPersonalityPersonaProfile} to dynamic systems with internal states and emotions~\cite{WangChiu2023HumanoidAgents,Feng2025EmotionallyIntelligent}, though challenges remain including systematic biases~\cite{Li2025LLMGeneratedPersona} and personality generation difficulties~\cite{Molchanova2025ExploringPersonality}. Ahmad et al.~\cite{Ahmad2025SimulatingUserDiversity} emphasize behavioral diversity in user simulation, while Sun et al.~\cite{Sun2022MetaphoricalUserSimulators} explore metaphorical approaches. Xie et al.~\cite{Xie2024HumanSimulacra} demonstrate multi-agent cognitive mechanisms producing personified responses aligned with target characters, while Castricato et al.~\cite{Castricato2024PERSONA} and Ge et al.~\cite{Ge2024PersonaHub} procedurally generate diverse personas from demographic data. Chu et al.~\cite{Chu2024CohesiveConversations} highlight conversational coherence for maintaining persona consistency across multi-turn interactions.

\paragraph{Multi-Agent Orchestration and Coordination} Decomposing complex tasks into specialized agents has emerged as a powerful paradigm for managing system complexity. Lee et al.~\cite{Lee2024OrchestraLLM} and Zhang et al.~\cite{Zhang2025AgentOrchestra} demonstrate efficient orchestration through routing frameworks that strategically select between models, reducing computational costs by 50\% while improving performance. Dang et al.~\cite{Dang2025MultiAgentCollaboration} and Tran et al.~\cite{Tran2025MultiAgentCollaboration} introduce dynamic orchestration with centralized coordinators trained via reinforcement learning, evolving from static to adaptive structures. Bernard and Balog~\cite{Bernard2024FormalCharacterizationUserSimulation} formalize dialogue state and action spaces for conversational systems, while Balog and Zhai~\cite{Balog2025UserSimulationEraGenerativeAI} and Davidson et al.~\cite{Davidson2023UserSimulationLLM} emphasize combining LLMs with additional components to capture cognitive processes. Raza et al.~\cite{Raza2024TRiSM} introduce metrics like Component Synergy Score and Tool Utilization Efficacy for quantifying collaboration quality, while Shu et al.~\cite{Shu2024EffectiveMultiAgent} demonstrate 90\% goal success rates in multi-agent collaboration, highlighting the importance of structured protocols.

\paragraph{Cognitive Architectures and State Management} Cognitive science provides crucial insights for agent design. Sumers et al.~\cite{Sumers2024CoALA} propose CoALA, drawing from symbolic AI to organize agents with modular memory components and structured action spaces mirroring human cognitive processes. Hu and Ying~\cite{Hu2025UnifiedMindModel} present architectures based on Global Workspace Theory incorporating perception, planning, reasoning, memory, and motivation components. For dialogue systems, Niu et al.~\cite{Niu2024EnhancingDST} and Xu et al.~\cite{Xu2024ChainOfThoughtDST} use LLM-backed agents with chain-of-thought reasoning to generate annotated dialogues for state tracking, while Levi and Kadar~\cite{Levi2025IntellAgent} introduce graph-based modeling for multi-turn dialogues with policy constraints. Mehri et al.~\cite{Mehri2025GoalAlignment} emphasize goal alignment ensuring state tracking remains consistent with user objectives. These architectures emphasize separation between working memory (state tracking) and behavioral planning (motivation systems), validating specialized agent approaches for complex dialogue domains~\cite{Rastogi2019SchemaGuidedDialogue,Yi2024MultiTurnDialogueSurvey,Mo2024HierTOD}.

\paragraph{Synthetic Data Generation and Evaluation} Agent-based systems require sophisticated evaluation methodologies beyond traditional metrics. Zhuge et al.~\cite{Zhuge2024AgentAsAJudge} show Agent-as-a-Judge achieves 90\% alignment with human consensus while reducing evaluation costs by 97\%, dramatically outperforming LLM-as-a-Judge approaches. For synthetic data generation, Suresh et al.~\cite{Suresh2024DiaSynth} use Chain of Thought reasoning to generate dialogues achieving 90.48\% of in-domain data performance, while Devanathan et al.~\cite{Devanathan2025WhySynthetic} introduce 18 linguistically grounded metrics revealing deficits in sentiment and behavioral realism. Evaluation frameworks must address task completion, output quality, consistency, and robustness~\cite{Mohammadi2025EvaluationBenchmarking}, with Zhu et al.~\cite{Zhu2025EvolutionaryEvaluationLLMAgents,Zhu2025AgenticBenchmarks} emphasizing preventing trivial shortcuts and ensuring agents genuinely leverage persona and state understanding. Wang et al.~\cite{Wang2025SurveyLLMAgents} identify critical limitations in role-playing, alignment, and knowledge boundaries that multi-agent approaches can address.
\section{Methodology}

Our methodology employs a three-agent architecture comprising a User Agent, State Tracking Agent, and Message Attributes Generation Agent (subsections \ref{subsec:user_agent}-\ref{subsec:msg_attr_gen}) that collaborate through structured protocols and strict behavioral rules. The system operates under defined constraints and conversation rules (subsection \ref{subsec:protocol}) to ensure reliable simulation with persona adherence and task completion accuracy.

\begin{figure}[h]
\centering
\includegraphics[width=\textwidth]{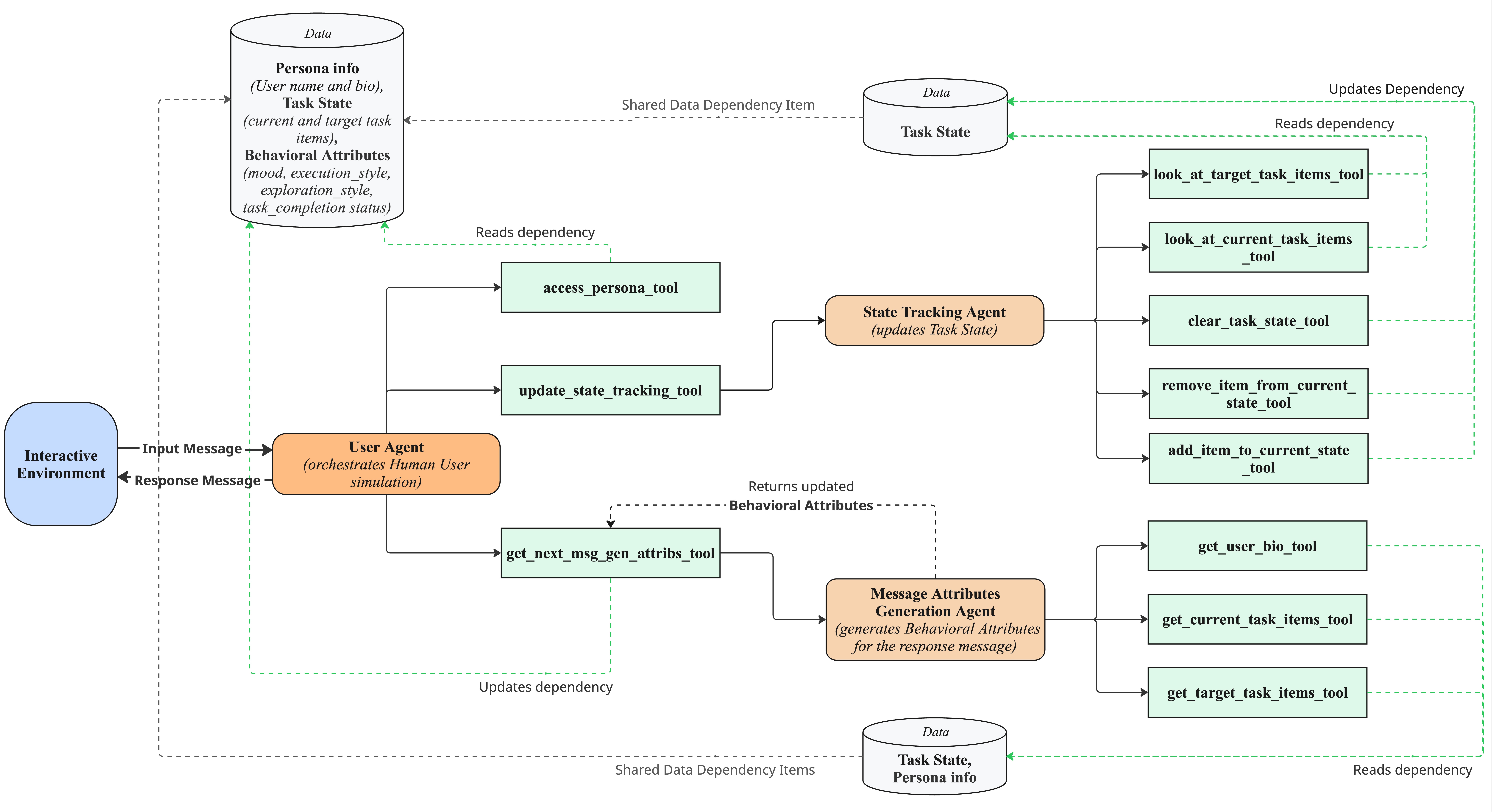}
\caption{\textbf{Multi-agent architecture for human user simulation} showing the three-agent framework: (1) \textbf{User Agent} serves as the primary orchestrator that generates simulated user responses by receiving input messages and invoking the two sub-agents in sequence, (2) \textbf{State Tracking Agent} maintains structured task state representation by tracking current confirmed items against target goals, and (3) \textbf{Message Attributes Generation Agent} determines behavioral characteristics (mood, execution style, exploration patterns) based on persona biography and current state.}
\label{fig:architecture}
\end{figure}

\subsection{User Agent}
\label{subsec:user_agent}

The User Agent serves as the primary orchestrator responsible for generating simulated user responses in the  conversation~\cite{Zhang2025AgentOrchestra,Dang2025MultiAgentCollaboration}. It receives the input messages and generates contextually appropriate responses to achieve task completion, using tool calls to fetch persona info and invoke the State Tracking and Message Attributes Generation agents as needed.

The user agent's response is generated as:
\begin{equation}
    r_t = f_{user}(m_t, s_t, a_t)
\end{equation}
where $r_t$ is the response at turn $t$, $m_t$ is the input message, $s_t$ is the task state from the State Tracking Agent,and $a_t$ is the behavioral attributes from the Message Attributes Generation Agent.

\subsection{State Tracking Agent}
\label{subsec:state_tracking}

The State Tracking Agent maintains a structured representation of the current state by parsing input messages to identify task items confirmed towards achieving the target state~\cite{Xu2024ChainOfThoughtDST,Mehri2025GoalAlignment}. The agent maintains two critical data structures:
\begin{itemize}
    \item $\mathcal{T}_{current}$   : A list of confirmed task items towards achieving the target task state
    \item $\mathcal{T}_{target}$    : The desired final task state that the user aims to achieve.
\end{itemize}

This agent uses its tools to add, remove or clear task items in the state. It updates the task state at turn $t$, as:
\begin{equation}
    s_t = f_{stateTracking}(input\_message) = \{\mathcal{T}_{current}, \mathcal{T}_{target}\} 
\end{equation}

\subsection{Message Attributes Generation Agent}
\label{subsec:msg_attr_gen}

The Message Attributes Generation Agent determines the behavioral characteristics for each user response~\cite{Feng2025EmotionallyIntelligent,WangChiu2023HumanoidAgents}, while using its tool to access the persona biography and the current task state. It outputs a structured set of behavioral attributes, $a_t$:
\begin{equation}
    a_t = \{mood\_tone, task\_execution_\_style, exploration\_style, task\_completion\_status\}
\end{equation}

\begin{itemize}
    \item $mood\_tone$ $\in$ \{casual, frustrated, confused, enthusiastic\}
    \item $task\_execution\_style$ $\in$ \{one-by-one, all-at-once\}
    \item $exploration\_style$ $\in$ \{explores, does-not-explore\}
    \item $task\_completion\_status$ $\in$ \{complete, incomplete\}
\end{itemize}

This agent's decisions are conditioned as:
\begin{equation}
    a_t = f_{msgAttrGen}(p_{bio}, s_t)
\end{equation}
where $p_{bio}$ is the persona biography, and $s_t$ is the current task state.

\subsection{Protocol}
\label{subsec:protocol}

\paragraph{Baseline instructions}

Each agent operates with specialized system instructions that define its role and constraints. The User Agent receives instructions to maintain persona consistency while working toward task completion. The State Tracking Agent focuses solely on accurate state extraction from input messages. The Message Attributes Generation Agent balances persona traits with appropriate behavioral variation.

\paragraph{Critical constraints}

To ensure reliable simulation, our agent instructions enforce critical constraints covering~\cite{Raza2024TRiSM}
\begin{enumerate}
    \item \textbf{Tool Invocation State}: The User Agent must invoke sub-agents in a specific sequence (State Tracking → Message Attributes Generation)~\cite{Tran2025MultiAgentCollaboration}
    \item \textbf{State Consistency}: State updates must be monotonic (task completion items are only added or removed, never implicitly modified, while keeping the execution within the bounds of $\mathcal{T}_{target}$)
    \item \textbf{Persona Boundaries}: Behavioral attributes must remain within persona-appropriate ranges
\end{enumerate}

\paragraph{Conversation rules}

The simulation follows structured conversation rules that govern the interaction flow: beginning with initial greeting and state intent expression, proceeding through progressive state building guided by the $task\_execution\_style$ attribute, handling clarification requests from the input, confirming state details before completion, and concluding with appropriate conversation closure once the stateing process is finished.

\paragraph{Exit gating}

The simulation terminates when the Message Attributes Generation Agent determines state completion ($task\_completion\_status = true$)~\cite{Zhang2025AgentOrchestra}. This decision is based on:
\begin{equation}
    task\_completion\_status = \begin{cases}
        true & \text{if } \mathcal{T}_{current} \supseteq \mathcal{T}_{target} \\
        false & \text{otherwise}
    \end{cases}
\end{equation}
\section{Experiments}

\subsection{Experimental Setup}

To validate our human user simulation framework, we implement and evaluate it in the domain of restaurant guest ordering. This domain presents an ideal testbed due to key characteristics aligning with human interaction complexities~\cite{Wang2020KddRES,Yi2024MultiTurnDialogueSurvey}. Restaurant ordering involves task complexity and ambiguity, requiring multi-turn conversations where guests navigate menu options, specify customizations, and handle clarifications~\cite{Mo2024HierTOD}. The ordering process incorporates complex state tracking as guests build orders with multiple items, each having various modifiers and customization options maintained throughout the conversation. The domain captures behavioral diversity through different foodie personas exhibiting distinct ordering styles, from methodical menu explorers to decisive quick-deciders, and varying emotional responses such as frustration with overwhelming choices or confusion about menu descriptions~\cite{Ge2024PersonaHub}. These persona-driven mood complexities, combined with the structured yet flexible ordering task, provide an excellent environment for testing our framework's ability to balance persona consistency, task completion accuracy, and realistic behavioral variation.

\paragraph{Datasets}

\begin{itemize}
    \item \textbf{Personas}: 20 diverse restaurant guest personas with distinct personality traits, communication styles, and behavioral preferences~\cite{Castricato2024PERSONA,Ahmad2025SimulatingUserDiversity}
    \item \textbf{Menu}: A comprehensive restaurant menu containing 50+ items across categories with various customization options and modifiers
    \item \textbf{Order Test Cases}: 60 test cases generated by pairing each persona with 3 different target orders of varying complexity (simple, medium, and complex orders with increasing customization levels)
\end{itemize}

\paragraph{Ordering System}
We evaluate our guest simulation by making it interact with an LLM-based ordering system (GPT-4o~\cite{Hurst2024GPT4oSystemCard,Liu2023AgentBench,Shu2024EffectiveMultiAgent}) configured with restaurant-specific instructions and menu knowledge. The ordering system greets customers, processes natural language order requests, clarifies ambiguous requests and suggests menu items, confirms order details and handles modifications, and completes transactions with order summaries, mimicking a real restaurant environment. The ordering system operates independently of our guest simulation, receiving only the conversation history and generating responses without knowledge of the testcase information or guest system's internal state.

\paragraph{Agentic Simulation Implementation}

Our implementation leverages Pydantic AI~\cite{pydanticai} with GPT-4o for structured multi-agent development with type-safe tool definitions and dynamic data dependency injection. The multi-agent implementation consists of the following key components: a \textbf{Main User Agent} that orchestrates the Guest Agent with sub-agent access tools to fetch persona information and invoke sub-agents to update order state and generate behavioral attributes for the next message in the ordering conversation; \textbf{Sub Agents} including the Order Tracking Agent and Message Attributes Generation Agent with specialized tools to update order state and generate behavioral attributes as needed by the Guest Agent; \textbf{Data Models} comprising Pydantic Basemodel classes to hold data dependencies~\cite{pydanticai_docs} defining the order state and behavioral attribute data objects; and \textbf{Conversation Management} featuring turn-limited interactions with repetition detection, tracking for tool calls, latencies, token usage, and structured conversation logging for comprehensive evaluation. This implementation was run on a local machine with a 10-core Apple-M1-Max CPU with 32GB of RAM and 3.2GHz of clock speed.

\subsection{Simulation Evaluation Metrics}

We evaluate our multi-agent framework using five quantitative metrics designed to capture different aspects of simulation quality.

\paragraph{Persona Adherence Score (PAS)}
Measures how well the user maintains their assigned persona throughout the conversation~\cite{Saggar2025ScoreBeforeSpeak,Wakaki2024ComperDial}. For each user message $i$ in a conversation with $N$ messages.

\begin{equation}
PAS = \frac{1}{N} \sum_{i=1}^{N} MS_i
\end{equation}

where the message score $MS_i$ is computed as:
\begin{equation}
MS_i = \sum_{j=1}^{4} w_j \cdot C_j
\end{equation}

with equal weights $w_j = 0.25$ for each component:
\begin{itemize}
    \item $C_1$: Exploration style match (explores vs. does not explore)
    \item $C_2$: Mood tone alignment (casual, frustrated, confused, enthusiastic)
    \item $C_3$: Task execution style match (one-by-one vs. all-at-once)
    \item $C_4$: Task completion status agreement
\end{itemize}

Each component $C_j \in \{0, 1\}$ is computed based on exact match with expected persona attributes.

\paragraph{Behavioral Variance Score (BVS)}
Captures realistic fluctuations in behavior to ensure natural human-like variations. For each behavioral dimension $d \in \{task\_execution\_style, exploration\_style, mood\_tone\}$:

\begin{equation}
TR_d = \frac{1}{M-1} \sum_{i=2}^{M} \mathbb{I}(state_i^d \neq state_{i-1}^d)
\end{equation}

where $M$ is the number of behavioral states. The average transition rate is:
\begin{equation}
TR_{avg} = \frac{TR_{task\_execution\_style} + TR_{exploration\_style} + TR_{mood\_tone}}{3}
\end{equation}

BVS uses a piecewise linear scoring function peaking at 20\% transition rate since humans typically exhibit moderate variance~\cite{Park2024GenerativeAgentSimulations} :
\begin{equation}    
BVS = \begin{cases}
\frac{TR_{avg}}{0.2} & \text{if } TR_{avg} \leq 0.2 \\
1 - \frac{TR_{avg} - 0.2}{0.8} & \text{if } TR_{avg} > 0.2
\end{cases}
\end{equation}

The range of BVS is [0, 1] where 1 is the best possible score. So we can expect robotic (too static) patterns having lower BVS scores while realistic patterns would have higher BVS scores.

\paragraph{Task Restriction Adherence (TRA)}
Evaluates accuracy in achieving the target state using F1-score with normalized task item matching~\cite{Jia2024LeveragingLLMs}. Task items are normalized by:
\begin{equation}
normalize(item) = lowercase(remove\_special(remove\_filler(item)))
\end{equation}

With $T$ = normalized target items and $C$ = normalized current state items,

\begin{equation}
Precision = \frac{|C \cap T|}{|C|}, \quad Recall = \frac{|C \cap T|}{|T|}
\end{equation}

\begin{equation}
TRA = \frac{2 \cdot Precision \cdot Recall}{Precision + Recall}
\end{equation}

\paragraph{Decision Explainability Index (DEI)}
Quantifies the traceability of the agentic system's internal decisions with tool usage results~\cite{Raza2024TRiSM}.

\begin{equation}
DEI = \begin{cases}
0 & \text{No tools (no explainability)} \\
\min(0.2, \frac{ED}{N} \times 0.2) & \text{Basic tools (about 20\% explainability)} \\
\min(0.5, \frac{ED}{N} \times 0.5) & \text{Basic tools + 1 subAgent (about 50\% explainability)} \\
\min(1.0, \frac{ED}{2N}) & \text{Full system (100\% explainability)}
\end{cases}
\end{equation}

where $ED$ is the count of explained decisions (tool invocations) across $N$ messages.

\paragraph{Composite Realism and Reliability Score (CRRS)}
Provides a unified score for overall user simulation quality using universal weights~\cite{Phy2020USLH}.

\begin{equation}
CRRS = 0.25 \cdot PAS + 0.20 \cdot BVS + 0.35 \cdot TRA + 0.20 \cdot DEI
\end{equation}

These weights reflect: TRA (35\%) as the primary task metric, PAS (25\%) for persona consistency, BVS (20\%) for naturalness, and DEI (20\%) for system validation. Note that DEI naturally adjusts based on the agentic system's experimental setup.

\subsection{Ablations}

We conduct systematic ablation studies across five experimental configurations to isolate the contribution of each component. 

\textbf{Config1 - Baseline LLM}: Single LLM with all information (persona, target order, conversation history) provided directly in the prompt. No agentic decomposition or tool use.

\textbf{Config2 - User Agent Only}: Guest Agent without sub-agents. Has direct access to persona, target order and conversation history through tools but no structured state tracking or behavioral control.

\textbf{Config3 - User Agent + State Tracking (ST) Agent}: Guest Agent with Order Tracking sub-agent. Maintains structured order state but lacks explicit behavioral control.

\textbf{Config4 - User Agent + Message Attributes Generation (MAG) Agent}: Guest Agent with Message Attributes Generation sub-agent. Has behavioral control but no structured state tracking.

\textbf{Config5 - Full System}: Complete multi-agent system with both Order Tracking and Message Attributes Generation sub-agents.


\subsection{Results}

We evaluate each configuration across five quantitative metrics defined in Section 3.3. All experiments are run with 60 test cases per configuration.

\begin{table}[h]
\centering
\begin{minipage}{0.48\textwidth}
\centering
\renewcommand{\arraystretch}{1.2}
\caption{Evaluation Metrics Across All Configurations}
\label{tab:comprehensive_metrics}
\begin{tabular}{lrrrrr}
\toprule
\textbf{Config} & \textbf{PAS} & \textbf{BVS} & \textbf{TRA} & \textbf{DEI} & \textbf{CRRS} \\
\midrule
1 & 0.589 & 0.218 & 0.608 & 0.000 & 0.404 \\
2 & 0.585 & 0.485 & 0.582 & 0.200 & 0.487 \\
3 & 0.554 & 0.689 & \textbf{0.785} & 0.498 & 0.651 \\
4 & \textbf{0.661} & 0.000 & 0.602 & 0.432 & 0.462 \\
5 & 0.706 & \textbf{0.839} & \textbf{0.785} & \textbf{0.994} & \textbf{0.818} \\
\bottomrule
\end{tabular}
\renewcommand{\arraystretch}{1.0}
\end{minipage}
\hfill
\begin{minipage}{0.38\textwidth}
\centering
\renewcommand{\arraystretch}{1.2}
\captionof{table}{Response Computation Costs}
\label{tab:computational_cost}
\begin{tabular}{lrr}
\toprule
\textbf{Config} & \textbf{Avg. Tokens} & \textbf{Avg. Latency(s)} \\
\midrule
1 & 6,618 & 5.08 \\
2 & 13,505 & 4.56 \\
3 & 24,580 & 36.30 \\
4 & 15,763 & 16.88 \\
5 & 14,789 & 23.16 \\
\bottomrule
\end{tabular}
\renewcommand{\arraystretch}{1.0}
\end{minipage}
\end{table}

Tables~\ref{tab:comprehensive_metrics} and \ref{tab:computational_cost} present the complete evaluation results and the computational costs incurred per response while executing each configuration.

\begin{minipage}{0.48\textwidth}
\centering
\renewcommand{\arraystretch}{1.2}
\captionof{table}{Statistical Significance: Full System vs Baseline}
\label{tab:statistical_significance}
\begin{tabular}{lrr}
\toprule
\textbf{Metric} & \textbf{p-value} & \textbf{Improve} \\
\midrule
PAS & 0.0037** & +19.9\% \\
BVS & 0.0000*** & +284.5\% \\
TRA & 0.0047** & +29.1\% \\
DEI & 0.0000*** & +100.0\% \\
CRRS & 0.0000*** & +102.6\% \\
\bottomrule
\multicolumn{3}{l}{** p < 0.01, *** p < 0.001} \\
\end{tabular}
\renewcommand{\arraystretch}{1.0}
\end{minipage}
\hfill
\begin{minipage}{0.48\textwidth}
\centering
\includegraphics[width=\textwidth]{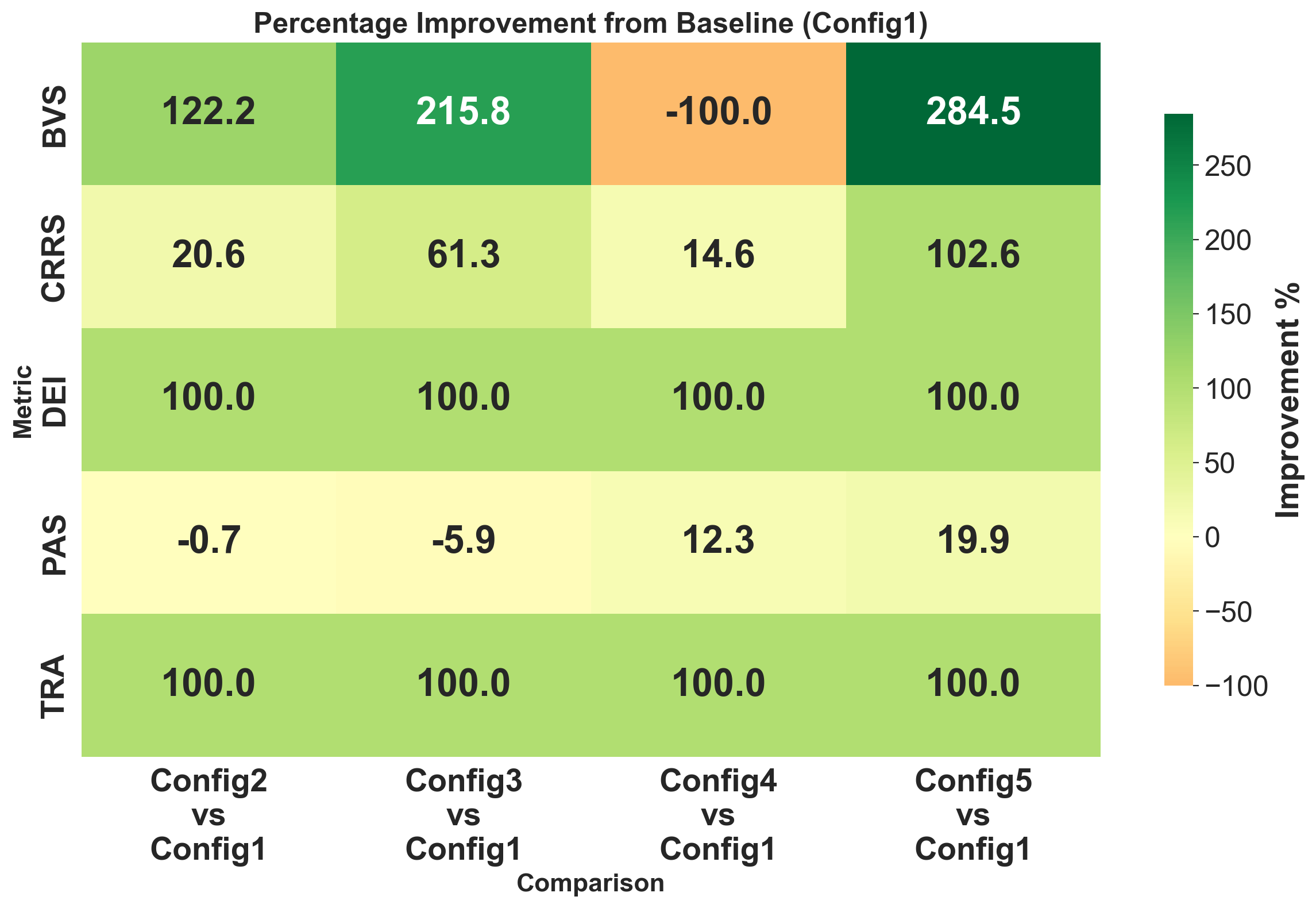}
\captionof{figure}{ Performance gains from baseline}
\label{fig:improvement_heatmap}
\end{minipage}

Table~\ref{tab:statistical_significance} presents the statistical significance analysis and Figure~\ref{fig:improvement_heatmap} visualizes the performance improvements across all configurations.

\subsubsection{Metrics Analysis}

\textbf{Persona Adherence Score (PAS)}: Config4 achieves the highest individual PAS (0.661), demonstrating the Message Attributes Agent's effectiveness for persona consistency. The full system (Config5) shows a 19.9\% improvement over baseline.

\textbf{Behavioral Variance Score (BVS)}: Config5 achieves the most realistic behavioral variance (0.839), representing a 284.5\% improvement over the static baseline. Notably, Config4 shows zero variance, indicating over-rigid behavioral control without order state awareness.

\textbf{Task Restriction Adherence (TRA)}: Both Config3 and Config5 achieve equivalent high performance (0.785), confirming the State Tracking Agent's critical role in maintaining order accuracy.

\textbf{Decision Explainability Index (DEI)}: Only Config5 achieves near-perfect explainability (0.994) through comprehensive tool usage traces, while simpler configurations lack decision transparency.

\textbf{Composite Realism and Reliability Score (CRRS)}: Config5 demonstrates the highest overall performance (0.818), achieving a \textbf{102.6\% simulation quality improvement} over baseline and outperforming all partial configurations.

\subsubsection{Key Findings}

\textbf{Component Synergy}: Neither sub-agent alone (Config3, Config4) achieves optimal performance; the combination (Config5) creates synergistic effects.

\textbf{Behavioral Rigidity}: Pure behavioral control (Config4) without state awareness leads to robotic, templated persona-default style interactions (BVS = 0).

\textbf{Cost-Performance Trade-off}: The full system balances computational cost (14,789 tokens) with superior performance across all metrics.

\textbf{Statistical Significance}: All improvements in Config5 over baseline are statistically significant (p < 0.01), indicating the effectiveness of our multi-agent architecture across all metrics for simulating human user behaviour in a task oriented conversation.

\section{Conclusion}

We presented a multi-agent framework decomposing human user simulation into specialized components: a User Agent for orchestration, State Tracking Agent for task management, and Message Attributes Generation Agent for behavioral control. Through persona control and task state grounding, our framework enables realistic simulation in conversational domains. Validating it in restaurant guest ordering scenarios demonstrates 102.6\% improvement in simulation quality over single-LLM baselines, with significant gains in persona adherence (19.9\%), behavioral variance (284.5\%), and task completion accuracy (29.1\%). These results validate that decomposing human simulation into specialized agents with coordinated state management yields superior performance across simulation quality dimensions, establishing our multi-agent architecture as effective for realistic human simulation in interactive AI systems.

\paragraph{Limitations and Ethical Considerations}
The multi-agent architecture incurs substantial computational costs (124\% more tokens, 356\% higher latency), limiting resource-constrained deployment. Each domain requires specialized prompt engineering and state design. Behavioral modeling lacks complex human behaviors like indecision, social dynamics, or cultural nuances~\cite{Wang2025SurveyLLMAgents}. Validation is English-only, single-domain with 60 test cases, restricting cross-cultural, multilingual, and real-world insights. Systems must maintain transparency and avoid bias perpetuation through auditing~\cite{Li2025LLMGeneratedPersona}. While democratizing conversational AI testing, the framework shouldn't impersonate individuals without consent or manipulate users believing they interact with real humans.

\paragraph{Future Work and Applications}
Key directions include adaptive persona evolution enabling dynamic behavioral adjustment~\cite{Cheng2024AutoPal}, multi-modal integration for voice-based prosodic features, visual gesture recognition, and emotional sentiment tracking~\cite{Chu2024MultimodalEmotional}, and cross-domain validation across customer support, e-commerce ordering~\cite{Xiang2024TransformerEcommerce}, healthcare consultations~\cite{Yu2024AIPatient}, educational tutoring~\cite{Maity2024GenerativeAI_ITS}, financial advisory, travel booking, and technical troubleshooting. Efficiency optimization through agent caching, selective tool invocation, and smaller specialized models could address computational overhead. The core principle of decomposing human behavioral simulation into specialized agents managing persona consistency and task state provides a foundation for realistic user simulations across interactive domains. Our framework offers systematic high-quality interaction data generation for testing, evaluation, and quality assurance wherever validating human-system interactions is critical for system reliability and user experience.

\section*{Acknowledgements}

Thanks to Andrew Norfleet (Principal Data Scientist at Toast) for his invaluable support during the ideation of the agentic simulation piece, and Benjamin Tang (Director of AI at Toast) for his support in pursuing the development of this work. This work was funded by Toast Inc.

{
    \small
    \bibliographystyle{plain}
    \bibliography{main}

@inproceedings{Zhuge2024AgentAsAJudge,
  title     = {{Agent-as-a-Judge: Evaluate Agents with Agents}},
  author    = {Mingchen Zhuge and Changsheng Zhao and Dylan R. Ashley and Wenyi Wang and Dmitrii Khizbullin and Yunyang Xiong and Zechun Liu and Ernie Chang and Raghuraman Krishnamoorthi and Yuandong Tian and Yangyang Shi and Vikas Chandra and Jürgen Schmidhuber},
  booktitle = {arXiv preprint arXiv:2410.10934},
  year      = {2024}
}

@techreport{Park2025SimulatingHumanBehavior,
  title        = {Simulating Human Behavior with AI Agents},
  author       = {Joon Sung Park and Carolyn Q. Zou and Aaron Shaw and Benjamin Mako Hill and Carrie J. Cai and Meredith Ringel Morris and Robb Willer and Percy Liang and Michael S. Bernstein},
  institution  = {Stanford Institute for Human-Centered Artificial Intelligence (HAI)},
  year         = {2025},
  month        = may,
  type         = {Policy Brief},
  url          = {https://hai.stanford.edu/assets/files/hai-policy-brief-simulating-human-behavior-with-ai-agents.pdf},
}

@inproceedings{Sutcliffe2023SurveyPersonalityPersonaProfile,
  title     = {{A Survey of Personality, Persona, and Profile in Conversational Agents and Chatbots}},
  author    = {Richard Sutcliffe},
  booktitle = {arXiv preprint arXiv:2401.00609},
  year      = {2023}
}

@inproceedings{WangChiu2023HumanoidAgents,
  title     = {{Humanoid Agents: Platform for Simulating Human-like Generative Agents}},
  author    = {Zhilin Wang and Yu Ying Chiu and Yu Cheung Chiu},
  booktitle = {Proceedings of the 2023 Conference on Empirical Methods in Natural Language Processing: System Demonstrations},
  year      = {2023},
  pages     = {167--176},
  address   = {Singapore},
  publisher = {Association for Computational Linguistics}
}

@inproceedings{Bernard2024FormalCharacterizationUserSimulation,
  title     = {{Towards a Formal Characterization of User Simulation Objectives in Conversational Information Access}},
  author    = {Nolwenn Bernard and Krisztian Balog},
  booktitle = {arXiv preprint arXiv:2406.19007v1},
  year      = {2024}
}

@inproceedings{Balog2025UserSimulationEraGenerativeAI,
  title     = {{User Simulation in the Era of Generative AI: User Modeling, Synthetic Data Generation, and System Evaluation}},
  author    = {Krisztian Balog and ChengXiang Zhai},
  booktitle = {arXiv preprint arXiv:2501.04410v1},
  year      = {2025}
}

@inproceedings{Wang2025SurveyLLMAgents,
  title     = {{A Survey on Large Language Model based Autonomous Agents}},
  author    = {Lei Wang and Chen Ma and Xueyang Feng and Zeyu Zhang and Hao Yang and Jingsen Zhang and Zhi-Yuan Chen and Jiakai Tang and Xu Chen and Yankai Lin and Wayne Xin Zhao and Zhewei Wei and Ji-Rong Wen},
  booktitle = {arXiv preprint arXiv:2308.11432},
  year      = {2025}
}

@inproceedings{Li2025LLMGeneratedPersona,
  title     = {{LLM Generated Persona is a Promise with a Catch}},
  author    = {Ang Li and Haozhe Chen and Hongseok Namkoong and Tianyi Peng},
  booktitle = {arXiv preprint arXiv:2503.16527v1},
  year      = {2025}
}

@inproceedings{Zhu2025EvolutionaryEvaluationLLMAgents,
  title     = {{Evolutionary Perspectives on the Evaluation of LLM-Based AI Agents: A Comprehensive Survey}},
  author    = {Jiachen Zhu and Menghui Zhu and Renting Rui and Rong Shan and Congmin Zheng and Bo Chen and Yunjia Xi and Jianghao Lin and Weiwen Liu and Ruiming Tang and Yong Yu and Weinan Zhang},
  booktitle = {arXiv preprint arXiv:2506.11102v1},
  year      = {2025}
}

@inproceedings{Zhu2025AgenticBenchmarks,
  title     = {{Establishing Best Practices for Building Rigorous Agentic Benchmarks}},
  author    = {Yuxuan Zhu and Tengjun Jin and Yada Pruksachatkun and Andy Zhang and Shu Liu and Sasha Cui and Sayash Kapoor and Shayne Longpre and Kevin Meng and Rebecca Weiss and Fazl Barez and Rahul Gupta and Jwala Dhamala and Jacob Merizian and Mario Giulianelli and Harry Coppock and Cozmin Ududec and Jasjeet Sekhon and Jacob Steinhardt and Antony Kellerman and Sarah Schwettmann and Matei Zaharia and Ion Stoica and Percy Liang and Daniel Kang},
  booktitle = {arXiv preprint arXiv:2507.02825v2},
  year      = {2025}
}

@inproceedings{Mohammadi2025EvaluationBenchmarking,
  title     = {{Evaluation and Benchmarking of LLM Agents: A Survey}},
  author    = {Mahmoud Mohammadi and Yipeng Li and Jane Lo and Wendy Yip},
  booktitle = {arXiv preprint arXiv:2507.21504v1},
  year      = {2025}
}

@inproceedings{Lee2024OrchestraLLM,
  title     = {{OrchestraLLM: Efficient Orchestration of Language Models for Dialogue State Tracking}},
  author    = {Chia-Hsuan Lee and Hao Cheng and Mari Ostendorf},
  booktitle = {arXiv preprint arXiv:2311.09758},
  year      = {2024}
}

@inproceedings{Park2024GenerativeAgentSimulations,
  title     = {{Generative Agent Simulations of 1,000 People}},
  author    = {Joon Sung Park and Carolyn Q. Zou and Aaron Shaw and Benjamin Mako Hill and Carrie Cai and Meredith Ringel Morris and Robb Willer and Percy Liang and Michael S. Bernstein},
  booktitle = {arXiv preprint arXiv:2411.10109},
  year      = {2024}
}

@inproceedings{Suresh2024DiaSynth,
  title     = {{DiaSynth: Synthetic Dialogue Generation Framework for Low Resource Dialogue Applications}},
  author    = {Sathya Krishnan Suresh and Wu Mengjun and Tushar Pranav and Eng Siong Chng},
  booktitle = {Proceedings of NAACL 2025},
  year      = {2025}
}

@inproceedings{Levi2025IntellAgent,
  title     = {{IntellAgent: A Multi-Agent Framework for Evaluating Conversational AI Systems}},
  author    = {Elad Levi and Ilan Kadar},
  booktitle = {arXiv preprint arXiv:2501.11067},
  year      = {2025}
}

@inproceedings{Sumers2024CoALA,
  title     = {{Cognitive Architectures for Language Agents}},
  author    = {Theodore R. Sumers and Shunyu Yao and Karthik Narasimhan and Thomas L. Griffiths},
  booktitle = {arXiv preprint arXiv:2309.02427},
  year      = {2024}
}

@inproceedings{Dang2025MultiAgentCollaboration,
  title     = {{Multi-Agent Collaboration via Evolving Orchestration}},
  author    = {Yufan Dang and Chen Qian and Xueheng Luo and others},
  booktitle = {arXiv preprint arXiv:2505.19591},
  year      = {2025}
}

@inproceedings{Hu2025UnifiedMindModel,
  title     = {{Unified Mind Model: Reimagining Autonomous Agents in the LLM Era}},
  author    = {Pengbo Hu and Xiang Ying},
  booktitle = {arXiv preprint arXiv:2503.03459},
  year      = {2025}
}

@inproceedings{Xie2024HumanSimulacra,
  title     = {{Human Simulacra: Benchmarking the Personification of Large Language Models}},
  author    = {Qiuejie Xie and Qiming Feng and Tianqi Zhang and others},
  booktitle = {Proceedings of ICLR 2025},
  year      = {2025}
}

@inproceedings{Raza2024TRiSM,
  title     = {{TRiSM for Agentic AI: A Review of Trust, Risk, and Security Management in LLM-based Agentic Multi-Agent Systems}},
  author    = {Shaina Raza and Ranjan Sapkota and Manoj Karkee and Christos Emmanouilidis},
  booktitle = {arXiv preprint arXiv:2506.04133},
  year      = {2024}
}

@inproceedings{Niu2024EnhancingDST,
  title     = {{Enhancing Dialogue State Tracking Models through LLM-backed User-Agents Simulation}},
  author    = {Cheng Niu and Xingguang Wang and Xuxin Cheng and Juntong Song and Tong Zhang},
  booktitle = {arXiv preprint arXiv:2405.13037},
  year      = {2024}
}

@inproceedings{Park2023GenerativeAgents,
  title     = {{Generative Agents: Interactive Simulacra of Human Behavior}},
  author    = {Joon Sung Park and Joseph C. O'Brien and Carrie J. Cai and Meredith Ringel Morris and Percy Liang and Michael S. Bernstein},
  booktitle = {arXiv preprint arXiv:2304.03442},
  year      = {2023}
}

@inproceedings{Castricato2024PERSONA,
  title     = {{PERSONA: A Reproducible Testbed for Pluralistic Alignment}},
  author    = {Louis Castricato and Nathan Lile and Rafael Rafailov and Jan-Philipp Fränken and Chelsea Finn},
  booktitle = {arXiv preprint arXiv:2407.17387},
  year      = {2024}
}

@inproceedings{Devanathan2025WhySynthetic,
  title     = {{Why Synthetic Isn't Real Yet: A Diagnostic Framework for Contact Center Dialogue Generation}},
  author    = {Rishikesh Devanathan and Varun Nathan and Ayush Kumar},
  booktitle = {arXiv preprint arXiv:2508.18210},
  year      = {2025}
}

@inproceedings{Molchanova2025ExploringPersonality,
  title     = {{Exploring the Potential of Large Language Models to Simulate Personality}},
  author    = {Maria Molchanova and Anna Mikhailova and Anna Korzanova and Lidiia Ostyakova and Alexandra Dolidze},
  booktitle = {Workshop on Customizable NLP (CustomNLP4U) at EMNLP 2024},
  year      = {2025}
}

@inproceedings{Ahmad2025SimulatingUserDiversity,
  title     = {{Simulating User Diversity in Task-Oriented Dialogue Systems using Large Language Models}},
  author    = {Adnan Ahmad and Stefan Hillmann and Sebastian Möller},
  booktitle = {arXiv preprint arXiv:2502.12813},
  year      = {2025}
}

@inproceedings{Chu2024CohesiveConversations,
  title     = {{Cohesive Conversations: Enhancing Authenticity in Multi-Agent Simulated Dialogues}},
  author    = {KuanChao Chu and Yi-Pei Chen and Hideki Nakayama},
  booktitle = {arXiv preprint arXiv:2407.09897},
  year      = {2024}
}

@inproceedings{Sun2022MetaphoricalUserSimulators,
  title     = {{Metaphorical User Simulators for Evaluating Task-oriented Dialogue Systems}},
  author    = {Weiwei Sun and Shuyu Guo and Shuo Zhang and Pengjie Ren and Zhumin Chen and Maarten de Rijke and Zhaochun Ren},
  booktitle = {arXiv preprint arXiv:2204.00763},
  year      = {2022}
}

@inproceedings{Mehri2025GoalAlignment,
  title     = {{Goal Alignment in LLM-Based User Simulators for Conversational AI}},
  author    = {Shuhaib Mehri and Xiaocheng Yang and Takyoung Kim and Gokhan Tur and Shikib Mehri and Dilek Hakkani-Tür},
  booktitle = {arXiv preprint arXiv:2507.20152},
  year      = {2025}
}

@inproceedings{Davidson2023UserSimulationLLM,
  title     = {{User Simulation with Large Language Models for Evaluating Task-Oriented Dialogue}},
  author    = {Sam Davidson and Salvatore Romeo and Raphael Shu and James Gung and Arshit Gupta and Saab Mansour and Yi Zhang},
  booktitle = {arXiv preprint arXiv:2309.13233},
  year      = {2023}
}

@inproceedings{Rastogi2019SchemaGuidedDialogue,
  title     = {{Towards Scalable Multi-domain Conversational Agents: The Schema-Guided Dialogue Dataset}},
  author    = {Abhinav Rastogi and Xiaoxue Zang and Srinivas Sunkara and Raghav Gupta and Pranav Khaitan},
  booktitle = {Proceedings of AAAI 2020},
  year      = {2020}
}

@inproceedings{Zhang2025AgentOrchestra,
  title     = {{AgentOrchestra: A Hierarchical Multi-Agent Framework for General-Purpose Task Solving}},
  author    = {Wentao Zhang and Liang Zeng and Yuzhen Xiao and others},
  booktitle = {arXiv preprint arXiv:2506.12508},
  year      = {2025}
}

@inproceedings{Xu2024ChainOfThoughtDST,
  title     = {{Chain of Thought Explanation for Dialogue State Tracking}},
  author    = {Lin Xu and Ningxin Peng and Daquan Zhou and See-Kiong Ng and Jinlan Fu},
  booktitle = {arXiv preprint arXiv:2403.04656},
  year      = {2024}
}

@inproceedings{Tran2025MultiAgentCollaboration,
  title     = {{Multi-Agent Collaboration Mechanisms: A Survey of LLMs}},
  author    = {Khanh-Tung Tran and Dung Dao and Minh-Duong Nguyen and Quoc-Viet Pham and Barry O'Sullivan and Hoang D. Nguyen},
  booktitle = {arXiv preprint arXiv:2501.06322},
  year      = {2025}
}

@inproceedings{Feng2025EmotionallyIntelligent,
  title     = {{Emotionally Intelligent Task-oriented Dialogue Systems: Architecture, Representation, and Optimisation}},
  author    = {Shutong Feng and Hsien-chin Lin and Nurul Lubis and Carel van Niekerk and Michael Heck and Benjamin Ruppik and Renato Vukovic and Milica Gašić},
  booktitle = {arXiv preprint arXiv:2507.01594},
  year      = {2025}
}

@inproceedings{Saggar2025ScoreBeforeSpeak,
  title     = {{Score Before You Speak: Improving Persona Consistency in Dialogue Generation using Response Quality Scores}},
  author    = {Arpita Saggar and Jonathan C. Darling and Vania Dimitrova and Duygu Sarikaya and David C. Hogg},
  booktitle = {Proceedings of ECAI 2025},
  year      = {2025}
}

@inproceedings{Phy2020USLH,
  title     = {{Deconstruct to Reconstruct a Configurable Evaluation Metric for Open-Domain Dialogue Systems}},
  author    = {Vitou Phy and Yang Zhao and Akiko Aizawa},
  booktitle = {Proceedings of COLING 2020},
  year      = {2020}
}

@inproceedings{Wakaki2024ComperDial,
  title     = {{ComperDial: Commonsense Persona-grounded Dialogue Dataset and Benchmark}},
  author    = {Hiromi Wakaki and Yuki Mitsufuji and Yoshinori Maeda and Yukiko Nishimura and Silin Gao and Mengjie Zhao and Keiichi Yamada and Antoine Bosselut},
  booktitle = {arXiv preprint arXiv:2406.11228},
  year      = {2024}
}

@inproceedings{Jia2024LeveragingLLMs,
  title     = {{Leveraging LLMs for Dialogue Quality Measurement}},
  author    = {Jinghan Jia and Abi Komma and Timothy Leffel and Xujun Peng and Ajay Nagesh and Tamer Soliman and Aram Galstyan and Anoop Kumar},
  booktitle = {arXiv preprint arXiv:2406.17304},
  year      = {2024}
}

@inproceedings{Wang2020KddRES,
  title     = {{KddRES: A Multi-level Knowledge-driven Dialogue Dataset for Restaurant Towards Customized Dialogue System}},
  author    = {Hongru Wang and Min Li and Zimo Zhou and Gabriel Pui Cheong Fung and Kam-Fai Wong},
  booktitle = {arXiv preprint arXiv:2011.08772},
  year      = {2020}
}

@inproceedings{Yi2024MultiTurnDialogueSurvey,
  title     = {{A Survey on Recent Advances in LLM-Based Multi-turn Dialogue Systems}},
  author    = {Zihao Yi and Jiarui Ouyang and Zhe Xu and Yuwen Liu and Tianhao Liao and Haohao Luo and Ying Shen},
  booktitle = {arXiv preprint arXiv:2402.18013},
  year      = {2024}
}

@inproceedings{Mo2024HierTOD,
  title     = {{HierTOD: A Task-Oriented Dialogue System Driven by Hierarchical Goals}},
  author    = {Lingbo Mo and Shun Jiang and Akash Maharaj and Bernard Hishamunda and Yunyao Li},
  booktitle = {arXiv preprint arXiv:2411.07152},
  year      = {2024}
}

@inproceedings{Ge2024PersonaHub,
  title     = {{Scaling Synthetic Data Creation with 1,000,000,000 Personas}},
  author    = {Tao Ge and Xin Chan and Xiaoyang Wang and Dian Yu and Haitao Mi and Dong Yu},
  booktitle = {arXiv preprint arXiv:2406.20094},
  year      = {2024}
}

@inproceedings{Cheng2024AutoPal,
  title     = {{AutoPal: Autonomous Adaptation to Users for Personal AI Companionship}},
  author    = {Yi Cheng and Wenge Liu and Kaishuai Xu and Wenjun Hou and Yi Ouyang and Chak Tou Leong and Wenjie Li and Xian Wu and Yefeng Zheng},
  booktitle = {arXiv preprint arXiv:2406.13960},
  year      = {2024}
}

@inproceedings{Chu2024MultimodalEmotional,
  title     = {{Towards Multimodal Emotional Support Conversation Systems}},
  author    = {Yuqi Chu and Lizi Liao and Zhiyuan Zhou and Chong-Wah Ngo and Richang Hong},
  booktitle = {arXiv preprint arXiv:2408.03650},
  year      = {2024}
}

@inproceedings{Yu2024AIPatient,
  title     = {{Simulated patient systems are intelligent when powered by large language model-based AI agents}},
  author    = {Huizi Yu and Jiayan Zhou and Lingyao Li and others},
  booktitle = {arXiv preprint arXiv:2409.18924},
  year      = {2024}
}

@inproceedings{Maity2024GenerativeAI_ITS,
  title     = {{Generative AI and Its Impact on Personalized Intelligent Tutoring Systems}},
  author    = {Subhankar Maity and Aniket Deroy},
  booktitle = {arXiv preprint arXiv:2410.10650},
  year      = {2024}
}

@inproceedings{Xiang2024TransformerEcommerce,
  title     = {{Text Understanding and Generation Using Transformer Models for Intelligent E-commerce Recommendations}},
  author    = {Yafei Xiang and Hanyi Yu and Yulu Gong and Shuning Huo and Mengran Zhu},
  booktitle = {arXiv preprint arXiv:2402.16035},
  year      = {2024}
}

@inproceedings{Hurst2024GPT4oSystemCard,
  title     = {{GPT-4o System Card}},
  author    = {Aaron Hurst and Adam Lerer and Adam P. Goucher and Adam Perelman and Aditya Ramesh and Aidan Clark and AJ Ostrow and Akila Welihinda and Alan Hayes and Alec Radford and Aleksander Mądry and Alex Baker-Whitcomb and Alex Beutel and Alex Borzunov and Alex Carney and Alex Chow and Alex Kirillov and Alex Nichol and Alex Paino and Alex Renzin and Alex Tachard Passos and Alexander Kirillov and Alexi Christakis and Alexis Conneau and Ali Kamali and others},
  booktitle = {arXiv preprint arXiv:2410.21276},
  year      = {2024}
}

@inproceedings{Liu2023AgentBench,
  title     = {{AgentBench: Evaluating LLMs as Agents}},
  author    = {Xiao Liu and Hao Yu and Hanchen Zhang and Yifan Xu and Xuanyu Lei and Hanyu Lai and Yu Gu and Hangliang Ding and Kaiwen Men and Kejuan Yang and Shudan Zhang and Xiang Deng and Aohan Zeng and Zhengxiao Du and Chenhui Zhang and Sheng Shen and Tianjun Zhang and Yu Su and Huan Sun and Minlie Huang and Yuxiao Dong and Jie Tang},
  booktitle = {arXiv preprint arXiv:2308.03688},
  year      = {2023}
}

@inproceedings{Shu2024EffectiveMultiAgent,
  title     = {{Towards Effective GenAI Multi-Agent Collaboration: Design and Evaluation for Enterprise Applications}},
  author    = {Raphael Shu and Nilaksh Das and Michelle Yuan and Monica Sunkara and Yi Zhang},
  booktitle = {arXiv preprint arXiv:2412.05449},
  year      = {2024}
}

@misc{pydanticai_docs,
  title        = {PydanticAI Documentation},
  howpublished = {\url{https://ai.pydantic.dev/}},
  note         = {Accessed: 2025-09-01},
  year         = {2024}
}

@misc{pydanticai,
  title        = {PydanticAI: A Python Agent Framework for LLMs},
  author       = {{Pydantic Team}},
  howpublished = {\url{https://github.com/pydantic/pydantic-ai}},
  note         = {Accessed: 2025-09-01},
  year         = {2024}
}
}

\end{document}